# Hydrogen and deuterium charging of site-specific specimen for atom probe tomography

Heena Khanchandani[1], Se-Ho Kim[1], Rama Srinivas Varanasi[1], TS Prithiv[1], Leigh T. Stephenson[1], Baptiste Gault[1,2]

[1]*Max-Planck-Institut für Eisenforschung, Max-Planck-Str. 1, 40237 Düsseldorf, Germany.*

[2]*Department of Materials, Royal School of Mines, Imperial College, Prince Consort Road, London SW7 2BP, United Kingdom.*

**Abstract:**

Hydrogen embrittlement can cause a dramatic deterioration of the mechanical properties of high-strength metallic materials. Despite decades of experimental and modelling studies, the exact underlying mechanisms behind hydrogen embrittlement remain elusive. To unlock understanding of the mechanism and thereby help mitigate the influence of hydrogen and the associated embrittlement, it is essential to examine the interactions of hydrogen with structural defects such as grain boundaries, dislocations and stacking faults. Atom probe tomography (APT) can, in principle, analyse hydrogen located specifically at such microstructural features but faces strong challenges when it comes to charging specimens with hydrogen or deuterium. Here, we describe three different workflows enabling hydrogen/deuterium charging of site-specific APT specimens: namely cathodic, plasma and gas charging. We discuss in detail the caveats of the different approaches in order to help with future research efforts and facilitate further studies of hydrogen in metals. Our study demonstrates successful cathodic and gas charging, with the latter being more promising for the analysis of the high-strength steels at the core of our work.

**Keywords:** atom probe tomography, hydrogen embrittlement, hydrogen trapping sites, twinning induced plasticity steel, cryogenic transfer workflows

## 1 Introduction

The degradation of the mechanical properties of metallic systems associated to the ingress of hydrogen leads to the premature catastrophic failures of many structural components [1–4]. A typical strategy to mitigate the deleterious influence of hydrogen on the material is hence to design alloys with a high number density of the trapping sites [5–7], which can act as irreversible traps, i.e. H is unable to re-enter the lattice under service conditions, owing to the high binding energy to such a site [8,9]. The trapped hydrogen could even potentially be beneficial in materials by increasing their resistance to hydrogen embrittlement [1,2,5]. In contrast to trapped hydrogen, diffusible hydrogen can diffuse through the material under ambient conditions [5], and, by interacting with crystalline defects, is believed to contribute to the deterioration of the mechanical properties [3,5,6,10]. In order to guide the design of hydrogen-resistant materials, it is necessary to study the details of the structure and composition of sites that can trap diffusible hydrogen, which are mostly defects such as stacking faults, dislocations and phase and grain boundaries [1,5]. Very few techniques have the combination of high spatial resolution and compositional sensitivity.

Atom probe tomography (APT) is a time-of-flight mass spectroscopy technique, which maps the spatial distribution of specific chemical species within a three-dimensional (3D) volume with sub-nanometre resolution [11,12]. In principle, APT is capable of detecting and quantifying hydrogen in three dimensions at near-atomic scale [13]. Yet despite some successes [14–17], and decades of work from numerous research groups, hydrogen microanalysis remains very challenging [1,2,13,14,17–20]. There are issues associated to the influence of residual gases from the analysis chamber of atom probe, specimen preparation and transport [20,21], and a strong dependence of the analytical performance on the analysis conditions [22–25]. Let us discuss these aspects in more details.

Studies have reported on the ionization of residual gases from the ultra-high vacuum analysis chamber of the atom probe, including hydrogen desorbed from the chamber walls, which can obscure the detection of hydrogen from the specimen [20,22,25,26]. To circumvent this issue, APT specimens are charged with deuterium (D or $^2$H) instead of hydrogen, as D has relatively lower natural abundance, i.e., 0.0156% of all hydrogen found on the earth. Residual hydrogen ionizes in the form of $H^+$, $H_2^+$ and $H_3^+$, and is therefore detected with characteristic peaks at 1, 2 and 3 Da respectively and their relative amplitude depends on the intensity of the electric field at the end of the field emitter [25,27]. The $H_2^+$ signal interferes with the D signal, making the quantification arduous [25]. It is hence crucial to minimize the likelihood of detecting $H_2^+$, which, typically, involves maximizing the strength of the electrostatic field. The use of high voltage pulsing rather than laser pulsing is hence recommended so as to minimize the influence of molecular ionic species of hydrogen [20,25,27].

Gemma *et al.* have studied the distribution of deuterium in Fe/V multi-layered thin films deposited on W [18,19]. They performed the APT experiments at different temperatures and demonstrated the influence of analysis temperature on D concentration profile [14]. Their studies also illustrated that even a small change in the local chemistry influences the D distribution substantially [14,18]. This work also exemplifies the impact of the analysis conditions on the quantitative deuterium distribution in materials, which was further evidenced in the study of bulk deuterides and hydrides [22–25].

Walck, Hren [1]; and Kellogg, Panitz [13] studied the trapping of implanted deuterium by structural defects such as vacancies induced by He implantation and a grain boundary in W. They found that deuterium, trapped by crystalline defects at cryogenic temperatures, diffuses and is subsequently released from the specimen as the temperature is raised. Hence, their studies [1,13] strongly suggest the necessity of cryogenic workflow according to which the charged specimen must be immediately quenched in order to retain the hydrogen into the reversible trapping sites of the specimen and further analysis on it must be carried out quickly [2,28]. This was further highlighted by more recent work reporting on hydrogen in steel [16,20]. These workflows involve the charging of APT tips with hydrogen or deuterium that are suitable for direct atom probe analysis [2,19,28].

However, these workflows were developed for wire-type samples prepared by electrochemical polishing, which do not allow for site-specific analyses. Breen et al. demonstrated that a substantial amount of hydrogen was introduced during the preparation of specimens by electrochemical polishing, leading to many of the trapping sites to be saturated and preventing effective deuterium charging [20]. Yet substantial progress in the field of APT has been achieved by the use of focused-ion beam (FIB) combined with scanning-electron microscopes (SEM) to prepare specimens from specific microstructural features [29]. As hydrogen interacts differently with different microstructural features, studying their trapping behavior to guide alloy design principles requires developing systematic workflows for charging the site-specific APT lift-outs with hydrogen or deuterium.

The current study describes our journey through three hydrogen/deuterium charging routes and their associated workflows: cathodic charging, plasma charging and gas charging suitable for charging specimens prepared by site-specific lift-out, as opposed to more conventional electropolished wires. We chose a twinning induced plasticity (TWIP) steel as our model system for the current study since it is highly susceptible to the hydrogen embrittlement [3,30]. The role of stacking faults, Σ3 twin boundaries and random boundaries has also been discussed in the literature with respect to their contribution to the hydrogen embrittlement resistance of TWIP steels [30–32]. Yet, the actual prevalent hydrogen embrittlement mechanism is not well understood. We discuss the details of these workflows with pros and cons, in order to help the community avoid some of the pitfalls associated to hydrogen and deuterium charging of APT specimens.

## 2  Experimental details
Materials
High manganese twinning induced plasticity (TWIP) steels contain over 20 wt.% Mn [33] and are austenitic, with a face centred cubic (FCC) crystal structure. A model TWIP steel with a chemical composition of Fe 28Mn 0.3C (wt.%) has been used for the current study. It was strip cast and subsequently homogenized at 1150°C for 2 hours. It was then 50% cold-rolled and recrystallized at 800°C for 20 minutes, followed by water cooling to room temperature.

Methods
A FEI Helios NanoLab 600i dual-beam FIB/SEM was used for preparing specimens for APT. APT experiments were conducted on either a LEAP 5000 XS or XR instrument (CAMECA Instruments Inc. Madison, WI, USA), in voltage pulsing mode at a set point temperature of 70K, 0.5% detection rate, 15-20% pulse fraction and 200kHz pulse repetition rate. These conditions had been defined in previous studies targeting an accurate detection of carbon in other austenitic steels [34].

For the cryogenic transfer workflows, we used the facilities developed within the framework of the Laplace project at the Max-Planck-Institute for Iron Research (MPIE), and detailed in ref [35]. This contains two atom probes, a CAMECA LEAP 5000 XS and a CAMECA LEAP 5000 XR, a FEI Helios Xe-plasma Focussed Ion Beam (PFIB), a $N_2$-atmosphere glovebox (Sylatec) and a gas-charging chamber known as the Reacthub Module [36]. These instruments are equipped with docking stations to host ultra-high vacuum carry transfer suitcases (UHVCTS, Ferrovac VSN40S) that are used for transferring the cryogenically cooled specimens between various instruments while minimizing contamination or frosting. The Xe-plasma FIB is also equipped with a cryogenically-cooled stage which allows cryo-FIB preparation of APT specimens and an intermediate chamber that enables the specimen transfers from the UHV suitcase to the cryo-stage. Cryogenic pucks were used to hold the APT specimens in the current study which are thermally insulated to avoid any direct contact with the vacuum transfer rods via a 2mm-thick layer of polyether ether ketone (PEEK).

# 3 Results

Figure 1 summarizes the seven workflows trialed, using three charging routes.

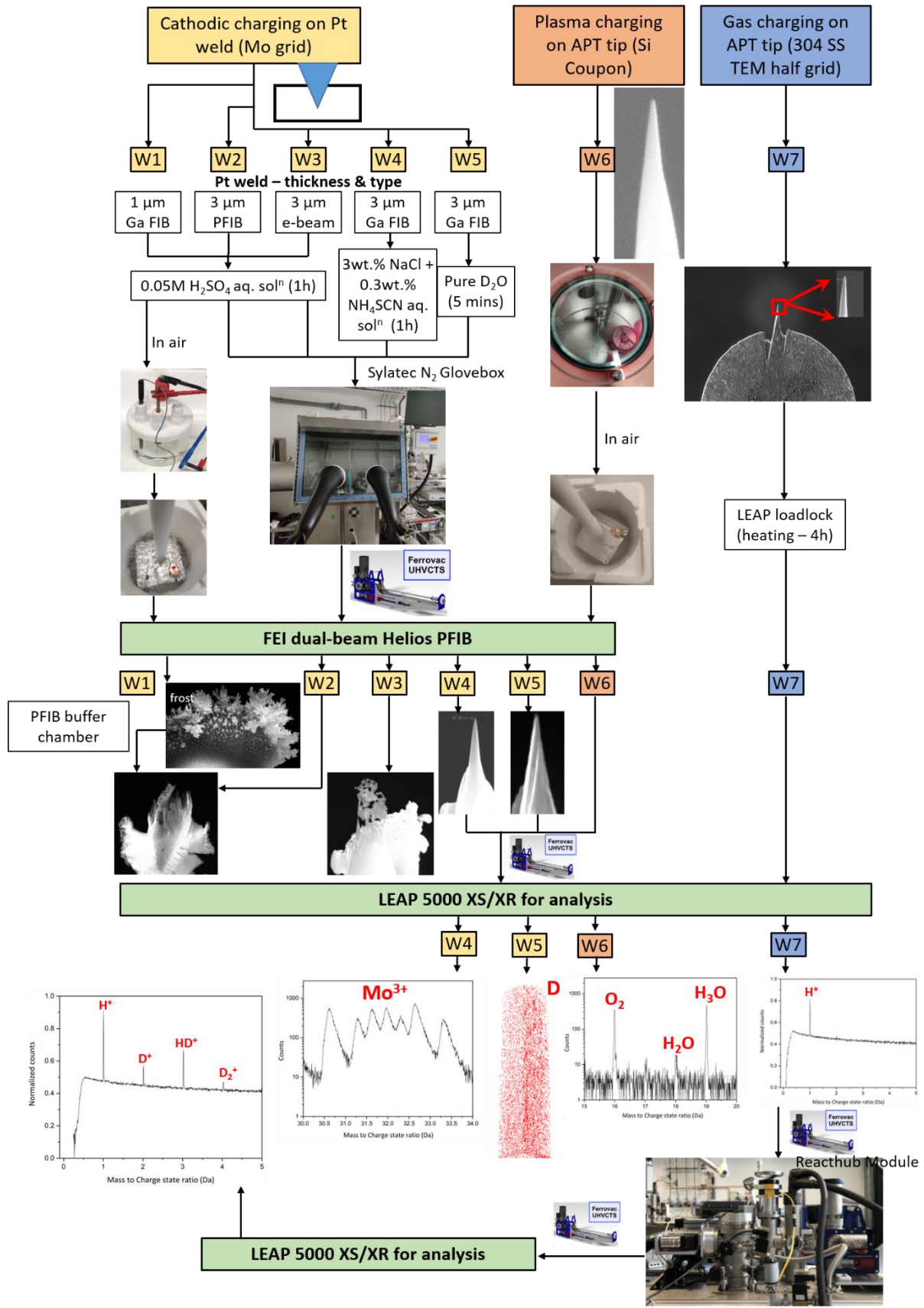

*Figure 1. Schematic depicting seven different workflows corresponding to three different charging routes: cathodic charging, plasma charging and gas charging.*

The detailed procedure of all workflows is described in subsequent sections.

### 3.1 Cathodic/electrolytic charging

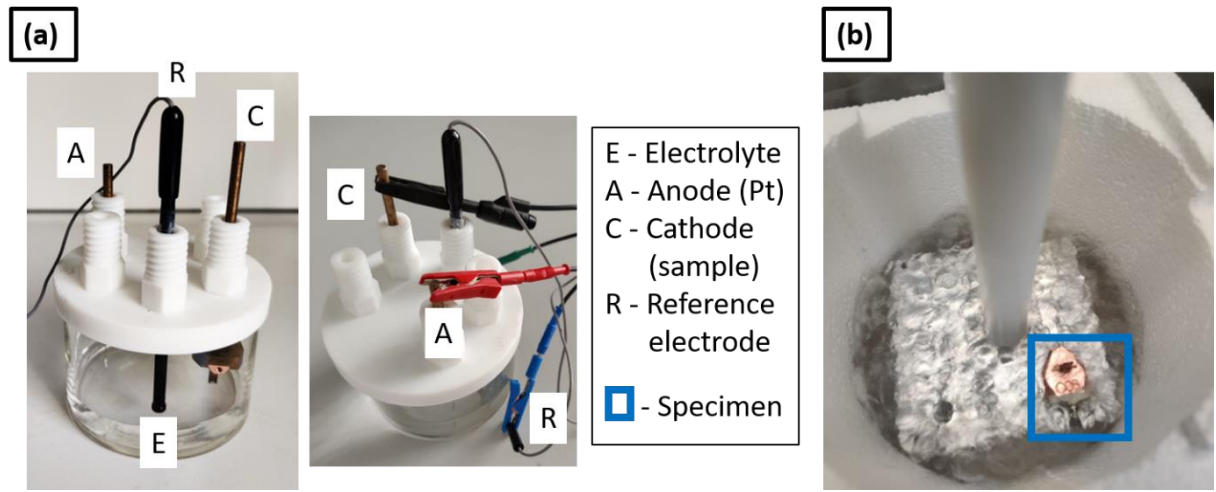

*Figure 2. (a) Set up for cathodic charging (electrolytic cell); (b) charged specimen inserted into an Al puck holder block immersed in LN$_2$ for an immediate quench.*

Cathodic/electrolytic charging of metallic specimens is carried out in an electrolytic cell, consisting of an electrolyte, i.e., the charging solution, the sample as cathode, a Pt wire as anode and a reference electrode connected to a power source. The electrolytic cell used in the present study is shown in Figure 2 (a) and was developed and used previously for wire-type samples prepared by electrochemical polishing – see ref. [20]. The process involves immersing the wire-specimen held inside the cryo-puck into the charging solution. Here, the aim was to develop the cathodic charging process for site-specific APT specimens mounted on a support. The conventional geometry used for APT site-specific specimens is a silicon coupon support held by a clip and mounted on a copper stub [29,37]. However Cu would dissolve in the solution, hindering the use of this conventional geometry.

Here, specimens were prepared on molybdenum grids which are typically used for TEM specimens and previously reported as a suitable alternative to support lift-outs for APT [38,39]. Figure 3 (a) shows an optical micrograph of the molybdenum grid with a diameter 3-mm which is cut using a razor blade into a half grid. The size of each Mo grid post is approx. 50-µm. The Mo grid is first electropolished in an aqueous solution of 10% NaOH (Sigma-Aldrich, Germany) by applying a DC voltage of ~ 10V [38–40]. After electropolishing, the size of Mo grid post is reduced to approx. 10-µm, as shown in the optical micrograph in Figure 3(b). This is then further sharpened by a Ga-FIB using annular milling process [29] in order to make the posts 2–3 µm in diameter, as visible from the SEM image in Figure 3 (c). The grid is held by a grid holder following the design detailed in ref. [38], and is used for mounting the samples for hydrogen charging. The grid holder is then inserted into the cryo-puck [35] which was fixed to the holder connected to the cathode of the electrolytic cell, shown in Figure 2 (a).

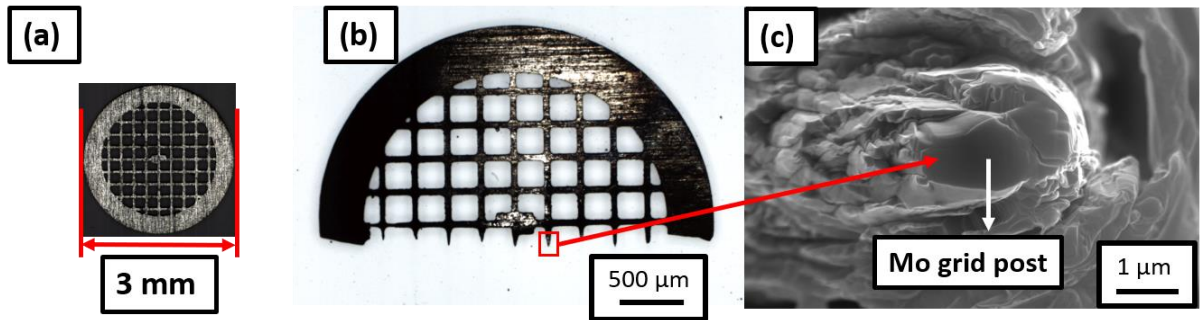

*Figure 3. (a) Molybdenum grid of diameter 3-mm; cut into (b) half grid; (c) Mo grid post of 2-3 microns in diameter suitable for mounting the samples.*

Site-specific APT specimen preparation was performed onto this Mo grid held in the grid holder following the protocol outlined in ref. [29]. The geometry is illustrated in Figure 4, with a schematic in Figure 4 (a). Here, "*t*" refers to the thickness of the Pt weld applied for mounting the specimen on the Mo grid post. Figure 4 (b) shows the SEM image of side view of the Pt weld obtained by tilting the grid holder at an angle of 90° while inserting it into the SEM chamber. Figures 4 (c) and (d) are scanning electron micrographs showing the top view of the Pt weld when the grid alignment is horizontal and vertical, respectively.

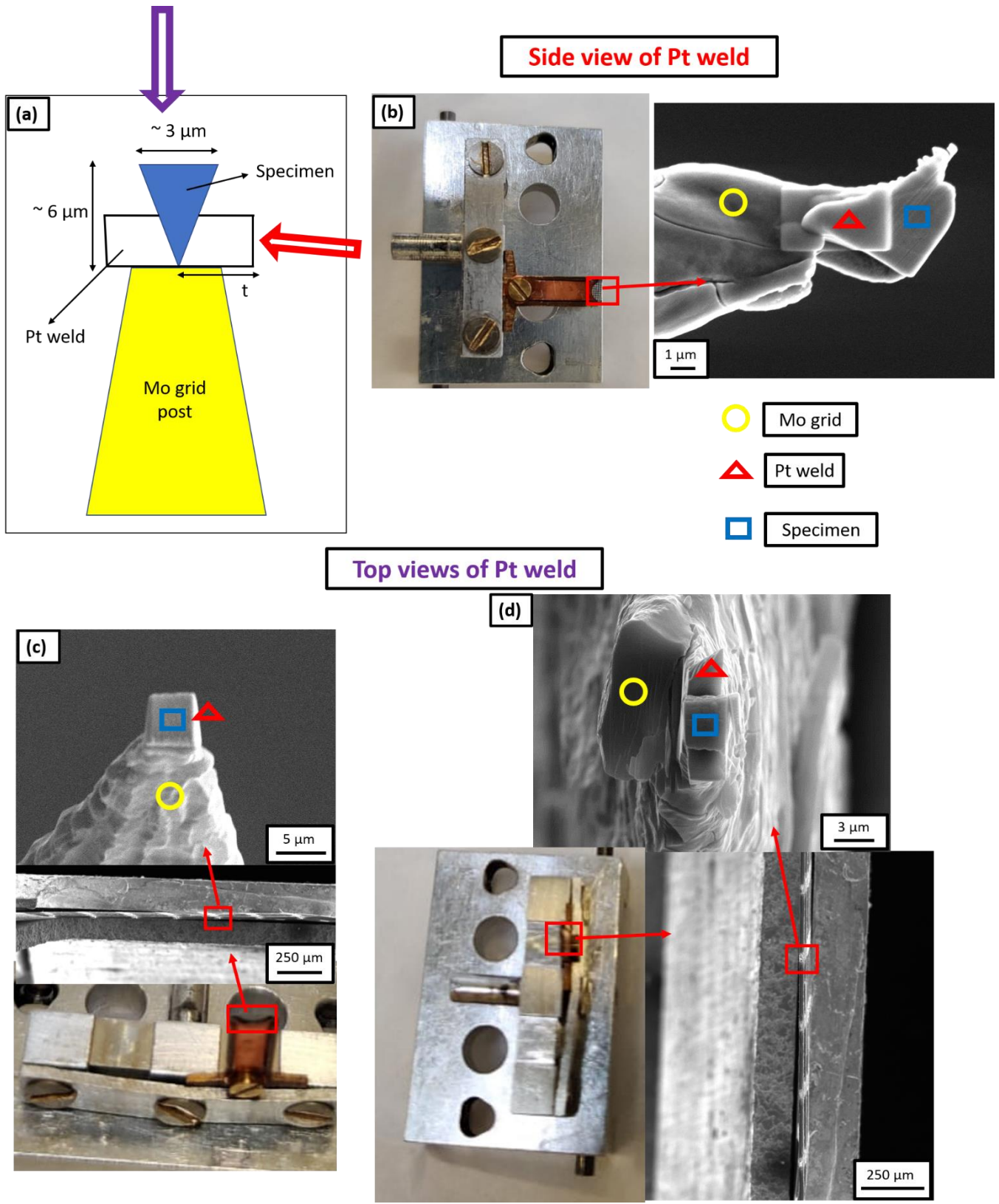

Figure 4. (a)Schematic of the Pt weld where "t" is the thickness of Pt weld; (b) SEM image of the side view of Pt weld obtained by tilting the grid holder at 90° while inserting it into the SEM chamber and the thickness of Pt weld is 3-µm in this case; (c) SEM image of the top view of Pt weld when the grid alignment is horizontal and the thickness of Pt weld is 1-µm in this case; (d) SEM image of the top view of Pt weld when the grid alignment is vertical and the thickness of Pt weld is 3-µm in this case.

*Workflow W1*

The Workflow W1 involves the preparation of a site-specific mounted lift-out using conventional Pt-welds, formed by decomposition of the Pt-precursor by using the Ga-ion beam, with a thickness of approx. 1-µm. The mounted lift-out is subjected to hydrogen charging prior to sharpening the deposited chunks into needle-shaped specimens. An aqueous solution of 0.05 M $H_2SO_4$ (Sigma-Aldrich, Germany, 98%) was used as an electrolyte for charging [41]. 1.4 g/l of thiourea (Merck Millipore, Germany) was mixed into the charging solution to act as hydrogen recombination binder [41]. It inhibits the combination of H atoms and prevents the formation of molecular $H_2$, thereby enhancing the hydrogen ingress into the specimen. A DC voltage of 1.5V was applied.

After 1 hour of charging at room temperature, with the charging cell on a bench in air, the sample was immediately inserted into an Al puck holder block immersed in liquid nitrogen ($LN_2$) for immediate quenching, as shown in Figure 2 (b). The Al block holding the specimen was transferred through ambient atmosphere into the cryogenically-cooled stage of the PFIB through an intermediate chamber that can be used as an air-lock, and not through the UHV suitcase. Upon pumping this intermediate chamber to high-vacuum conditions, the puck was inserted into the PFIB with the help of the dedicated transfer rod – see ref. [35] for details – in order to sharpen the specimen through annular milling [29]. Frost was observed on the Mo grid, as readily visible in Figure 5 (a). This could have been formed because the charging and transfers were performed in ambient atmosphere, which contains moisture. In this case, the sample was transferred back to the transfer rod of the PFIB [35] and allowed to stay on it for 5 minutes in order to desorb the frost under high vacuum of $10^{-7}$ mbar [28] and ambient temperature. The specimen was then transferred back to the cryostage for sharpening. Figure 5 (b) shows SEM image of the specimen during the sharpening, which indicates that the Pt-welds were strongly affected by the charging or the frosting.

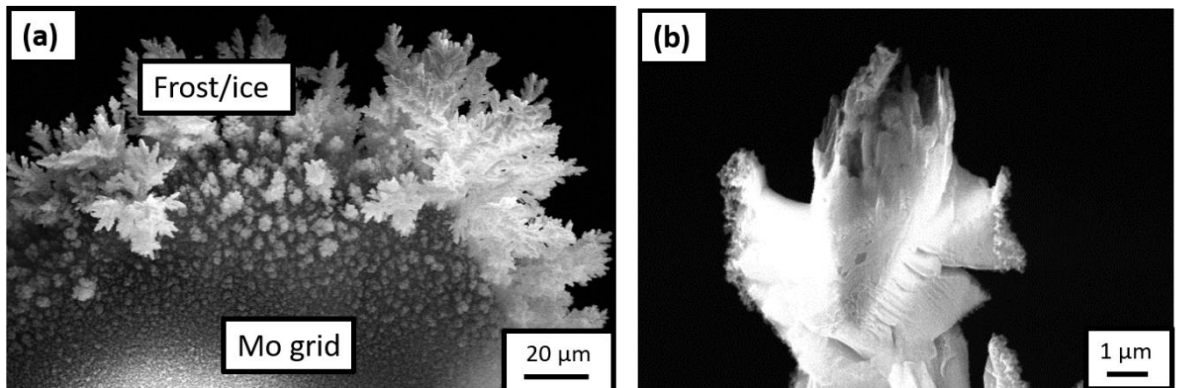

*Figure 5. (a) SEM image of the frost observed on the Mo grid after it was transferred onto the PFIB cryostage following the H-charging in an aqueous solution of 0.05M $H_2SO_4$ and quenching in air; (b) SEM image of the specimen during sharpening in the PFIB.*

*Workflow W2*

To avoid the formation of frost, all subsequent charging experiments were performed in a $N_2$ filled glovebox (Sylatec). The specimen, electrolytic cell, charging solution and liquid nitrogen ($LN_2$) bath with the Al puck holder were first loaded into the $N_2$ glovebox. The $LN_2$ bath inside the $N_2$ glovebox must be filled for quenching the sample immediately after charging [17]. After filling the $LN_2$ bath and precooling the Al block puck holder, the charging of the specimen was started in the electrolytic cell. After performing the charging and quenching in the $N_2$ atmosphere, the samples were transferred into the PFIB through the precooled UHV suitcase [35].

Specifically here in Worklow W2, we mounted the lifted-out chunk by using approx. 3-µm thick Pt-welds in the PFIB, formed by decomposition of the Pt-precursor using Xe-plasma FIB. The charging solution was the same as the one used in Workflow W1. Here again, the Pt-welds did not survive and their condition, after charging and quenching, was same as that of the previous experiment (Workflow W1), shown in Figure 5 (b).

*Workflow W3*

As a next step, in Workflow W3, since the Pt-welds appeared to be the weaker part of the mounted sample, we prepared a new batch of specimens by using e-beam induced Pt deposition to form larger and denser welds, with a thickness of approx. 3-µm. Subsequently, these specimens were charged following the same Workflow as previously. SEM imaging of the specimens in the PFIB after transfer through the precooled UHV suitcase revealed that the Pt-weld had survived but the sample had corroded, as shown in Figure 6. This experiment through Workflow W3 evidences that the steel sample corrodes in the acidic charging solution, leading to the dissolution of the specimen during charging. The duration of charging is also a critical factor to be considered because a longer charging duration could also lead to the dissolution of the specimen, beyond the issue of the weak Pt-welds.

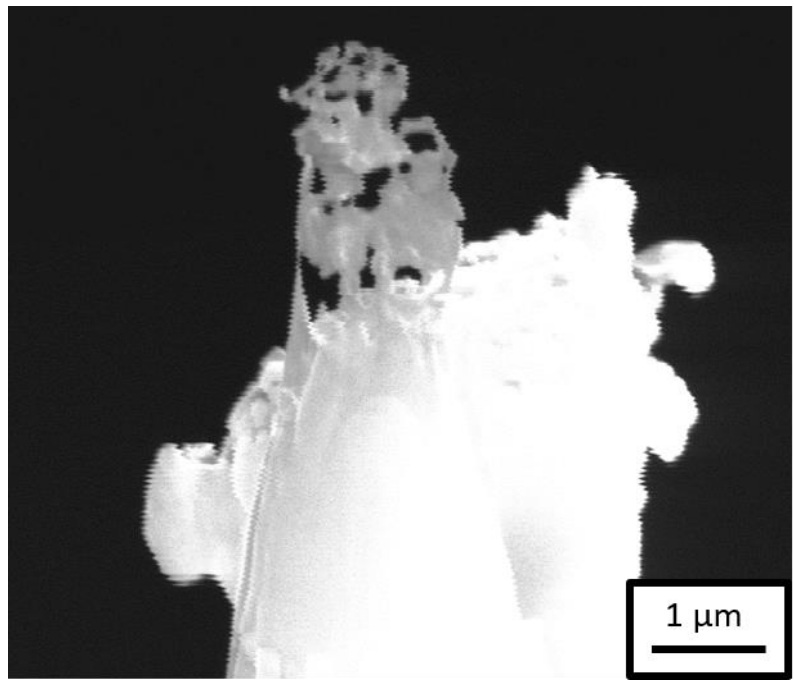

*Figure 6. SEM image of the specimen after H-charging in an aqueous solution of 0.05M $H_2SO_4$ and transferring through the precooled UHV suitcase during sharpening in the PFIB.*

*Workflow W4*

In the next experiment, Workflow W4, we decided to change the charging solution and use Ga-FIB induced Pt deposition for preparing the specimens with 3-µm thick Pt-welds. A neutral solution consisting of an aqueous solution of 3 wt.% NaCl (Sigma-Aldrich, Germany) with 0.3 wt.% $NH_4SCN$ (Sigma-Aldrich, Germany, 98%) as hydrogen recombination binder was used for charging [42]. The neutral charging solution is comparatively less effective for charging because only $H_2O$ is the source of $H^+$ ions in this solution. These $H^+$ ions move towards the cathode and penetrate into the sample which is connected to the cathode. $Na^+$ and $Cl^-$ ions only enhance the ionic conductivity of the solution, whereas $H_2SO_4$ was also the source of $H^+$ ions in the previously used acidic charging solution. Nevertheless, it should be sufficient for charging the APT specimens (Pt-welds) due to their extremely small, i.e., microscopic surface area.

After charging the specimen in the neutral charging solution for 1 hour, the sample and weld both survived, the sample did not corrode and it was sharpened successfully in the PFIB. Figure 7(a) is a SEM image of the sharpened specimen, which was taken to the LEAP 5000 XS for measurement through the precooled UHV suitcase. The corresponding mass spectrum is shown in Figure 7 (b) in which a substantial amount of Mo was observed. Mo could have originated from dissolution of the grid itself by the charging solution. At the applied voltage (1.5V) to the neutral electrolyte, Fe and Mo could have formed a galvanic couple due to which the following reaction can take place at the cathode:

$$H_2MoO_4 + 6H^+ + 3e^- \leftrightarrow Mo^{3+} + 4H_2O \qquad E° \rightarrow +0.43V$$

$Mo^{3+}$ from the solution would have deposited onto the sample which was observed in the mass spectrum[43].

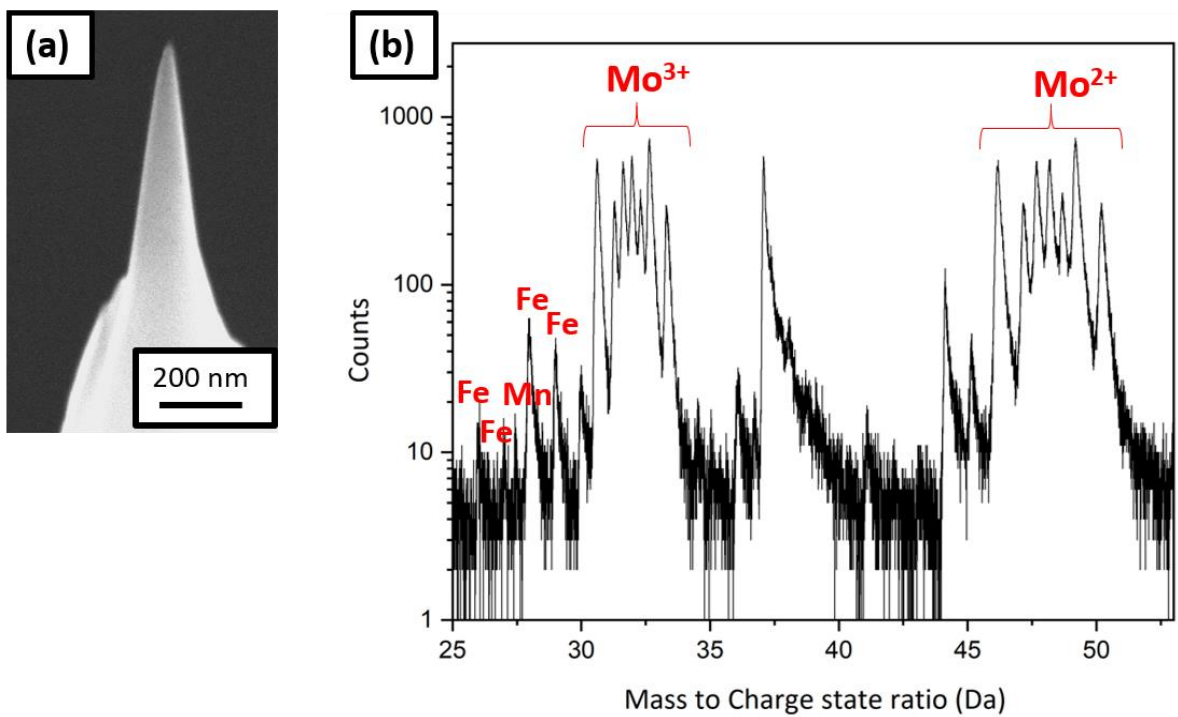

*Figure 7. (a) SEM image of the tip sharpened in the PFIB cryostage after charging in the neutral solution and quenching; (b) the corresponding mass spectrum showing Mo peaks.*

*Exchange of the charging solution*

Both the acidic and the alkaline solutions, typically used for electropolishing of Mo grid, must be avoided for charging the steel of interest. Consequently, it was decided to use pure $D_2O$ (Sigma-Aldrich, Germany, 99.9 %) for charging and shorter charging times ranging from 1 to 5 minutes.

Some pure $D_2O$ charging experiments were also performed on non-site-specific lift-outs sharpened APT specimens. These needle-shaped APT specimens were prepared from the steel sample by electropolishing [20] and then fine milled using the PFIB. The needle was charged in the $N_2$ glovebox in pure $D_2O$ for 1 minute. The applied voltage was 2.2V, to induce the dissociation of the $D_2O$ electrolyte into $D^+$ and $OD^-$ ions [17]. Then it was immediately quenched and transferred to the LEAP 5000 XS for measurement through the precooled UHV suitcase. Clusters of $H_2O$, $D_2O$ and associated complex ions were found in the mass spectrum consistent with the analysis of water by APT [44,45], but as soon as ions from the metal started to be detected, the specimen fractured.

For the next experiment, the sample was transferred to the PFIB immediately after charging and quenching in the $N_2$ glovebox through the precooled UHV suitcase. SEM shows frost on the charged needle (see Figure 8 (a)), associated to the freezing of the remaining water adsorbed on the specimen's surface. The specimen was therefore cleaned and resharpened in the PFIB post charging and quenching, Figure 8 (b), before being transferred to the LEAP 5000 XR for measurement.

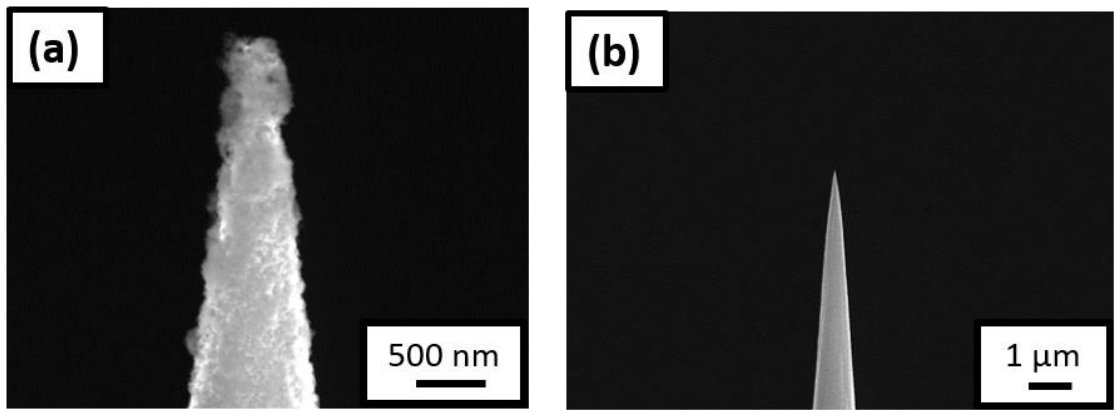

*Figure 8. (a) SEM image of the non-site specific lift-out specimen having the needle shaped geometry, surrounded by ice post charging and immediate quench; (b) SEM image of the same needle after being resharpened in the PFIB cryostage.*

In a subsequent experiment, we tried to clean the charged needle with ethanol immediately before quenching. The sample was then transferred to the LEAP 5000 XS for measurement directly. However, this induced a slight delay in the quenching process – albeit only by few seconds. The mass spectrum exhibited peaks likely associated to adsorbed ethanol frozen on the specimen's surface, and the specimen also fractured early.

*Workflow W5*

A set of pure $D_2O$ charging experiments was performed on a new batch of specimens prepared with approx. 3-µm thick Pt-welds formed by Ga-FIB induced Pt deposition. Multiple experiments were performed to optimize the charging time as charging for longer duration was leading to early specimen failure during measurement and charging for few seconds is not adequate to charge the sample with Pt-weld as the volume of material to charge is comparatively large compared to the needle-shaped geometry. Eventually, the charging performed for 5 minutes (Workflow W5) resulted in acceptable specimen yield. An SEM image of a sharpened specimen is shown in Figure 9 (a), and the corresponding 3D elemental map of carbon and deuterium are shown in Figures 9 (b) and (c) respectively following reconstruction.

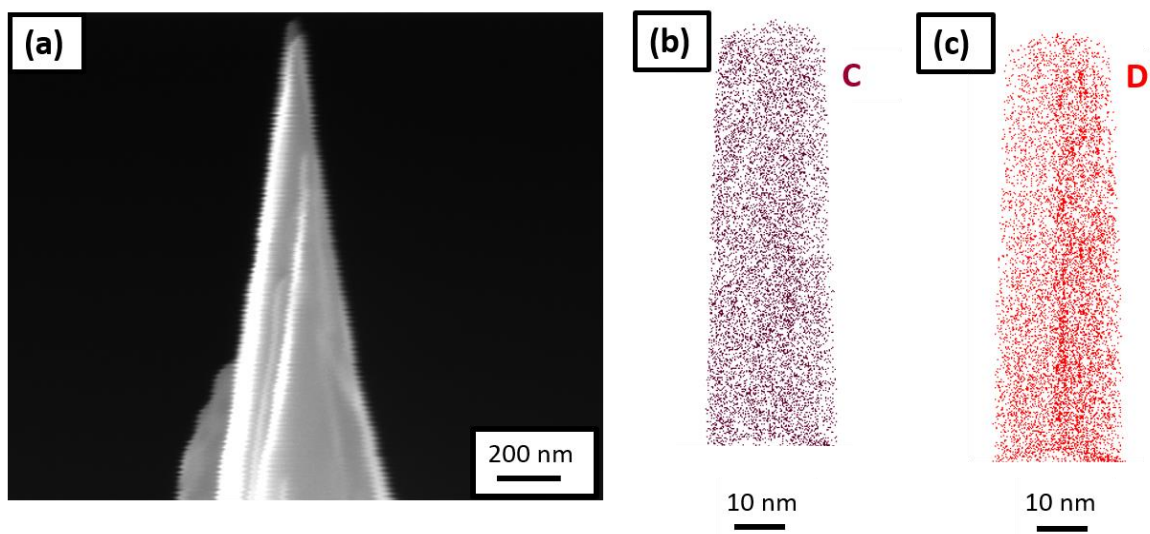

*Figure 9. (a) SEM image of the tip sharpened in the PFIB cryostage after charging the specimen in pure $D_2O$ and quenching; (b) 3D elemental map of carbon; and (c) deuterium, following its reconstruction.*

This final workflow demonstrates successful electrolytic charging of site-specific APT specimens. The (re)sharpening of the specimen in the PFIB following charging can, however, cause the introduction of frozen $D_2O$ from the surface because of implantation by the incoming high-energy ions [21]. $D_2O$ molecules could also dissociate and lead to D ingress into the sample during milling. These effects can be detrimental to the analysistical performance but can be minimized or avoided by using lower acceleration voltages and lower beam current during (re)sharpening for instance.

### 3.2 Plasma charging (*Workflow W6*)

A plasma cleaner (Evactron), shown in Figure 10 (a), was used for plasma charging. The plasma cleaner is normally used for removing carbon contamination from SEM samples with an oxygen plasma. Here the plasma cleaner was fed by a gas line of Ar containing 5% $H_2$ gas. For plasma charging, the samples were prepared conventionally on the silicon coupon held by a clip onto a copper stub, as the reactivity of copper in solution is no longer a concern.

Workflow W6 illustrates the plasma charging protocol. The APT specimens were prepared conventionally in the Ga FIB [37] and were charged by the hydrogen-rich plasma for 1 h (shown in Figure 10(b)) at a pressure of $10^{-1}$ mbar. Subsequently, they were immediately quenched in $LN_2$ as shown by Figure 10 (c), and were transferred into the PFIB, on the pre-cooled cryostage through the intermediate chamber as depicted by Figure 10 (d). The SEM micrograph in Figure 10 (e) shows no frost on the specimen's surface. It was resharpened at low kV (16 kV ion acceleration voltage and 30 pA beam current). Figure 10 (f) shows the SEM image of the tip after cleaning.

Finally, it was transferred into the LEAP 5000 XR for measurement through the precooled UHV suitcase and analysed. The corresponding mass spectrum is shown in Figure 10 (g) and eventually, the specimen fractured after only 100000 ions. During the APT analysis, hydrogen was detected in the mass spectrum at 1, 2 and 3 Da but its origin cannot be discerned. It could be from the specimen or from the analysis chamber of the atom probe. In addition, localised clusters of $H_2O$, $H_3O$ were observed. These could originate from the vacuum in the plasma cleaner, which cannot be below $10^{-2}$ mbar, and hence not devoid of oxygen and moisture that introduced inside the specimen along with the hydrogen during charging. It might also be from mild frosting after plunge freezing and the introduction of O-containing surface species with the Xe-beam in the PFIB during sharpening.

Due to the relatively poor vacuum and associated implantation of impurities, we did not pursue this route for many more attempts. In principle, plasma cleaner directly attached to some commercial instruments could also be directly used to implant H by changing the gas supply for instance, which would enable these implantations in a cleaner environment and avoid the transfer through the UHV suitcase. However this was, so far, not attempted.

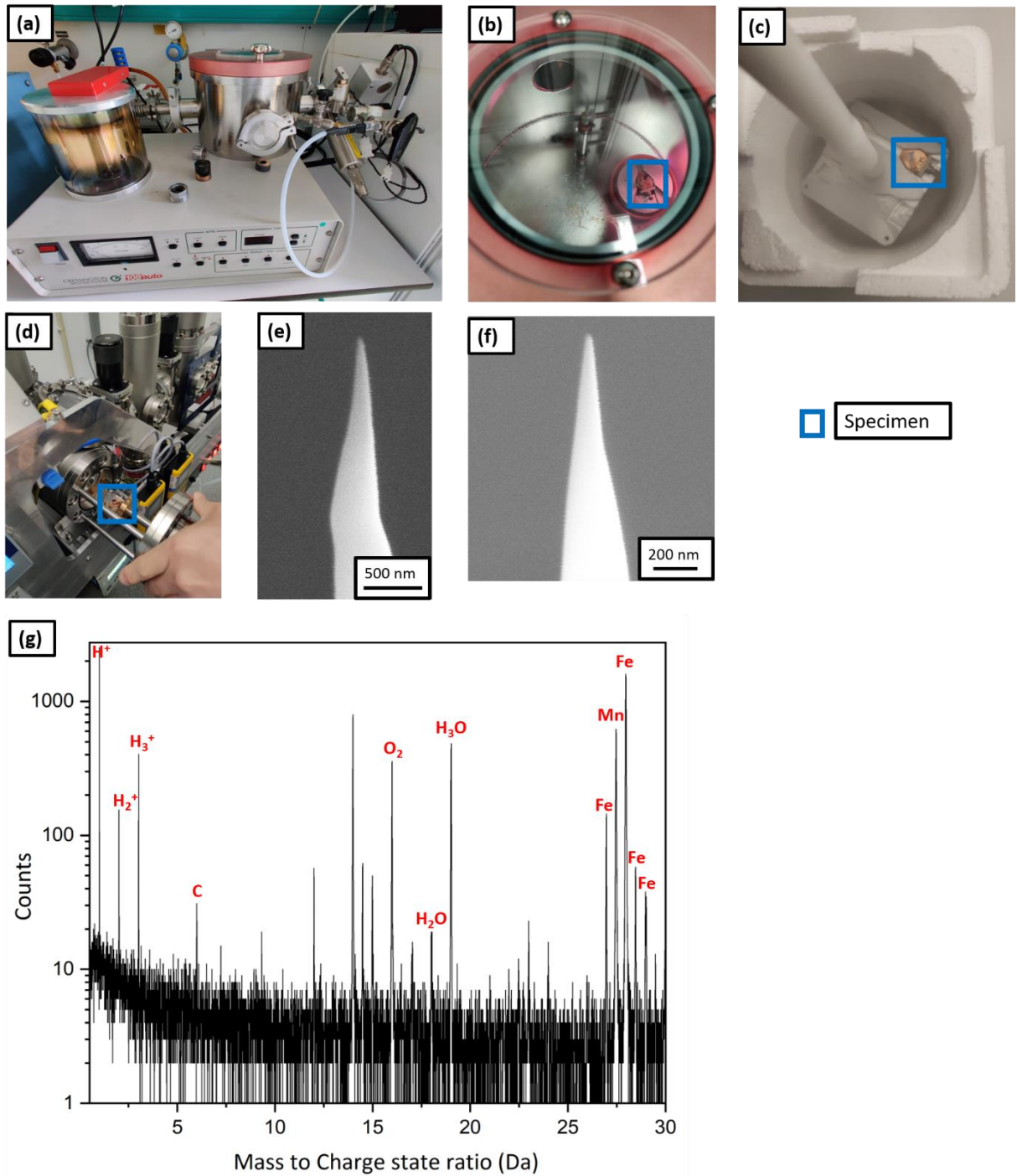

*Figure 10. (a) Plasma cleaner set up; (b) APT specimen being charged with the hydrogen plasma; (c)*

*the specimen placed on the precooled Al block for immediate quench post charging; (d) the specimen being transferred to the intermediate chamber of PFIB; (e) SEM image of the tip post charging and quenching; (f) SEM image of the same tip after resharpening in PFIB; (g) the corresponding mass spectrum.*

### 3.3 Gas charging (*Workflow W7*)

Gas charging was carried out in the RHM [36], as illustrated by Workflow W7. The RHM is equipped with a laser and hydrogen/deuterium gas flow so as to perform charging at a selected temperature while the sample is held on a cryo-stage to enable fast quenching. The samples are prepared on a cold-rolled stainless steel 304 (SS304) TEM half-grid, provided by JPT and CAMECA, as the RHM's pyrometer is calibrated for this steel grade. Site-specific APT specimens were prepared by using the lift-out procedure described above with the Ga FIB [37]. It was transferred into the LEAP 5000 XR loadlock where it was subjected to an outgassing heat treatment at 150°C and $10^{-7}$ mbar pressure for approximately 4 hours in order to desorb the hydrogen previously trapped within the material, as suggested in ref. [20]. This short heat treatment is not expected to substantially modify the microstructure but help desorb some of the trapped H within the specimen's microstructure arising from the initial specimen preparation.

The specimen was subsequently transferred into the analysis chamber and an APT analysis was started in order to clean the specimen's surface up to a voltage of 3–4 kV. The mass spectrum obtained from this pre-charging analysis shows only a peak at 1 Da, as shown in Figure 11 (a). It is expected that this hydrogen originates primarily from the residual gas of the analysis chamber, but could also in part be from the hydrogen trapped inside the material that was not desorbed by the outgassing heat treatment. Then it was transferred to the RHM for charging through the UHV suitcase. Deuterium gas charging was carried out in the RHM at a pressure of 250 mbar for 6 hours at 200°C, followed by an immediate quench down to 45K on the cryostage of the RHM. The specimen was then transferred back into the LEAP 5000 XR for further measurement through the precooled UHV suitcase. The post charging mass spectrum exhibits peaks at 2, 3 and 4 in addition to the one at 1 Da as shown in Figure 11 (b); which confirms the D charging of the specimen.

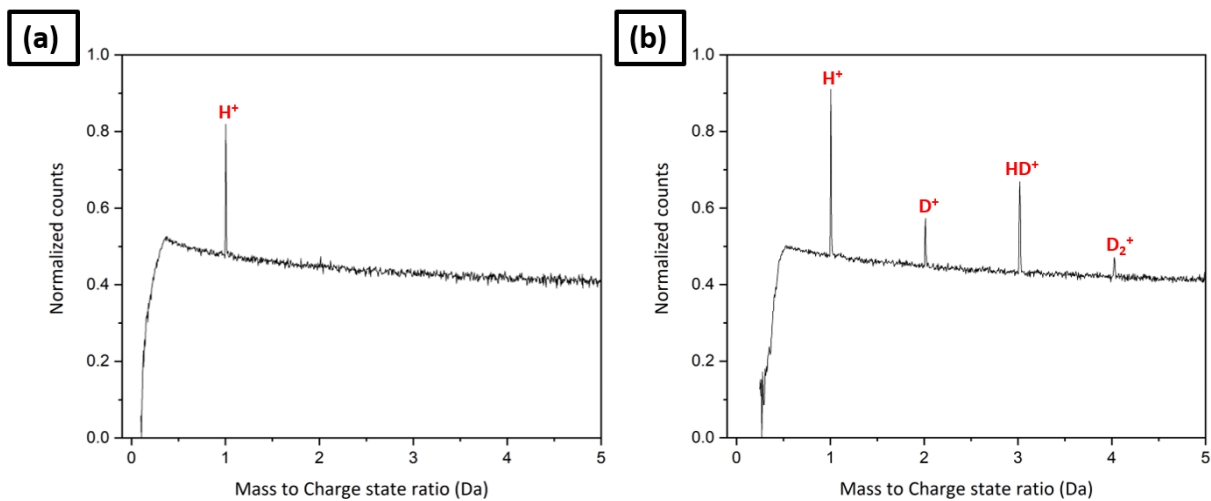

*Figure 11. (a) The pre-charging mass spectrum; (b) the mass spectrum after D gas charging in the RHM at a pressure of 250 mbar for 6 hours at 200°C.*

In Figure 11 (b), the peak at 1 Da has higher counts in the mass spectrum following charging than that of the tip before charging (Figure 11 (a)). Further, the Table 1 shows the bulk composition analysis in ionic %. H/Fe ratio is also higher in the post-charging mass spectrum than in the pre-charging mass spectrum. It implies that the residual background hydrogen is higher when the charged specimens are loaded into the analysis chamber. It could be misinterpreted as hydrogen originating from the specimen if the pre-charging mass spectrum had not been first collected and the intensity of the electric field is not compared between the pre- and post-charging measurements. The average charge state ratios of $Fe^+/Fe^{2+}$, corresponding to the pre- and post-charging mass spectra are comparable, as depicted in the Table 1, indicating towards the comparable electric field in both the analyses.

*Table 1. Bulk composition analysis (ionic %) corresponding to the pre- and post-charging mass spectra*

|  | **Pre-charging mass spectrum** | **Post-charging mass spectrum** |
|---|---|---|
| **Amount of H (ionic %)** | 1.436% | 3.193% |
| **Amount of D (ionic %)** | 0 | 0.1% |
| $\frac{H}{Fe}$ ratio | 0.019 | 0.049 |
| $\frac{Fe^+}{Fe^{2+}}$ ratio | 0.009 | 0.011 |

## 4 Discussion

Table 2 summarizes all the workflows corresponding to three different charging methods and their outcomes. The success of two workflows is evidenced from Table 2 - Workflow W5 through cathodic charging and Workflow W7 through gas charging.

*Table 2. Summary of three different charging routes*

| Charging method | Charging parameters | | | | | | Result |
|---|---|---|---|---|---|---|---|
| | *Work flow* | *Pt weld ('t': thickness)* | *Charging solution* | *Charging time* | *Voltage applied (V)* | *Cryogenic transfer atmosphere* | |
| | W1 | t - 1µm (Ga FIB) | 0.05M $H_2SO_4$ aq. sol$^n$ + 1.4 g/l thiourea | 1 hour | 1.5 | Air | ✖ |
| | W2 | t - 3µm (PFIB) | 0.05M $H_2SO_4$ aq. sol$^n$ + 1.4 g/l thiourea | 1 hour | 1.5 | $N_2$ Glovebox and UHV suitcase | ✖ |
| *Cathodic charging* | W3 | t - 3µm (e-beam) | 0.05M $H_2SO_4$ aq. sol$^n$ + 1.4 g/l thiourea | 1 hour | 1.5 | $N_2$ Glovebox and UHV suitcase | ✖ |

| | W4 | t - 3µm (Ga FIB) | 3 wt.% NaCl + 0.3 wt.% NH$_4$SCN aq. sol$^n$ | 1 hour | 1.5 | N$_2$ Glovebox and UHV suitcase | ✗ |
|---|---|---|---|---|---|---|---|
| | W5 | t - 3µm (Ga FIB) | Pure D$_2$O | 5 minutes | 2.2 | N$_2$ Glovebox and UHV suitcase | ✓ |
| **Plasma charging** | W6 | Ga FIB sharpened APT tip on Si coupon | Ar-5% H$_2$ plasma (10$^{-1}$ torr) | 1 hour | | Air | ✗ |
| **Gas charging** | W7 | Ga FIB sharpened APT tip on 304 SS TEM half grid | 250 mbar D gas | 6 hours | | UHV suitcase | ✓ |

In Workflow W5, the mounted lift-out chunk with approx. 3-µm thick Pt-weld prepared by Ga-FIB induced Pt deposition was cathodically charged in pure D$_2$O in N$_2$ glovebox for 5 minutes at an applied voltage of 2.2 V. It was then immediately quenched in LN$_2$ bath and transferred to the PFIB cryostage for sharpening through the precooled UHV suitcase. The sharpened specimen was transferred to the atom probe for measurement through the precooled UHV suitcase. Whilst Workflow W7 consists of deuterium gas charging of a sharpened APT tip in RHM at a pressure of 250 mbar for 6 hours at 200°C followed by an immediate quench down to 45 K on the cryostage of the RHM.

Gas charging enables data acquisition from the same specimen before and after charging, which is useful to distinguish between hydrogen coming from charging and the one which comes from the analysis chamber. This is not possible in the cathodic charging route, as the mounted lift-out chunk is charged before it is sharpened to enable APT analysis, so only post-charging data can be acquired. The pre-charging mass spectrum is essential for determining the amplitude of the background hydrogen signal, as had been pointed out by Walck and Hren [1]. They also collected several mass spectra from a Ni specimen prior to D-implantation. Also, no other specimen must be allowed to enter into the analysis chamber after the pre-run so as to avoid any change in the vacuum conditions of the analysis chamber due to the hydrogen coming from loading other samples.

However, since the detection of H as atomic or molecular ionic species depends on the strength of the electrostatic field [27], the relative H/D amounts across datasets can only be compared if they are acquired under similar electric field conditions. We estimated the electric field by calculating the average charge state ratios (CSR). We selected the peaks of C and CSR is computed as $^{12}C^{2+}/^{12}C^{1+}$ by using the peaks at m/q = 6 Da and m/q = 12 Da, containing $^{12}C^{2+}$ and $^{12}C^{1+}$ respectively. Admittedly the field evaporation behaviour can be more complex [46] and can be subject to additional detection issues [47,48], the carbon peaks were selected because they can be found across all datasets, conversely to Fe$^{1+}$ which was sometimes not detected above background. Multiple datasets of charged and uncharged specimens were used from both the LEAP 5000 XS and LEAP 5000 XR atom probes, for plotting the average charge state ratios vs. relative H abundances in Figure 12 (a) and (b), respectively.

Three relative H abundances are plotted in each: $H_1/H_{total}$, $H_2/H_{total}$ and $H_3/H_{total}$ where $H_{total}$ is $(H_1 + H_2 + H_3)$. $H_1$, $H_2$ and $H_3$ correspond to peaks obtained at 1, 2 and 3 Da respectively. The precision in the measurement of all the involved species is computed as their counting statistics [12]:

$$\sigma_i = \sqrt{\frac{C_i \times (1 - C_i)}{N}}$$

where $\sigma_i$ is the precision (counting statistics), $C_i$ is the atomic fraction of the element $i$, N is the number of measured ions which lie in between 100000 to $30 \times 10^6$ ions for the measurements used for plotting these curves.

The x-axis error bar corresponding to the CSR is computed as:

$$\sigma_{CSR} = \sqrt{(\sigma_{C^{2+}})^2 + (\sigma_{C^{1+}})^2}$$

The y-axis error bar corresponding to the three relative H abundances is computed as:

$$\sigma_{H_1/H_{total}}, \sigma_{H_2/H_{total}} \text{ and } \sigma_{H_3/H_{total}} = \sqrt{(\sigma_{H_1})^2 + (\sigma_{H_2})^2 + (\sigma_{H_3})^2}$$

All reported data from the LEAP 5000 XS are from cathodically charged specimens with hydrogen, except one H-gas charged in the RHM at room temperature at a pressure of 250 mbar. All reported from the LEAP 5000 XR are from D-gas charged specimens in the RHM except the one which was plasma charged (highlighted in light green).

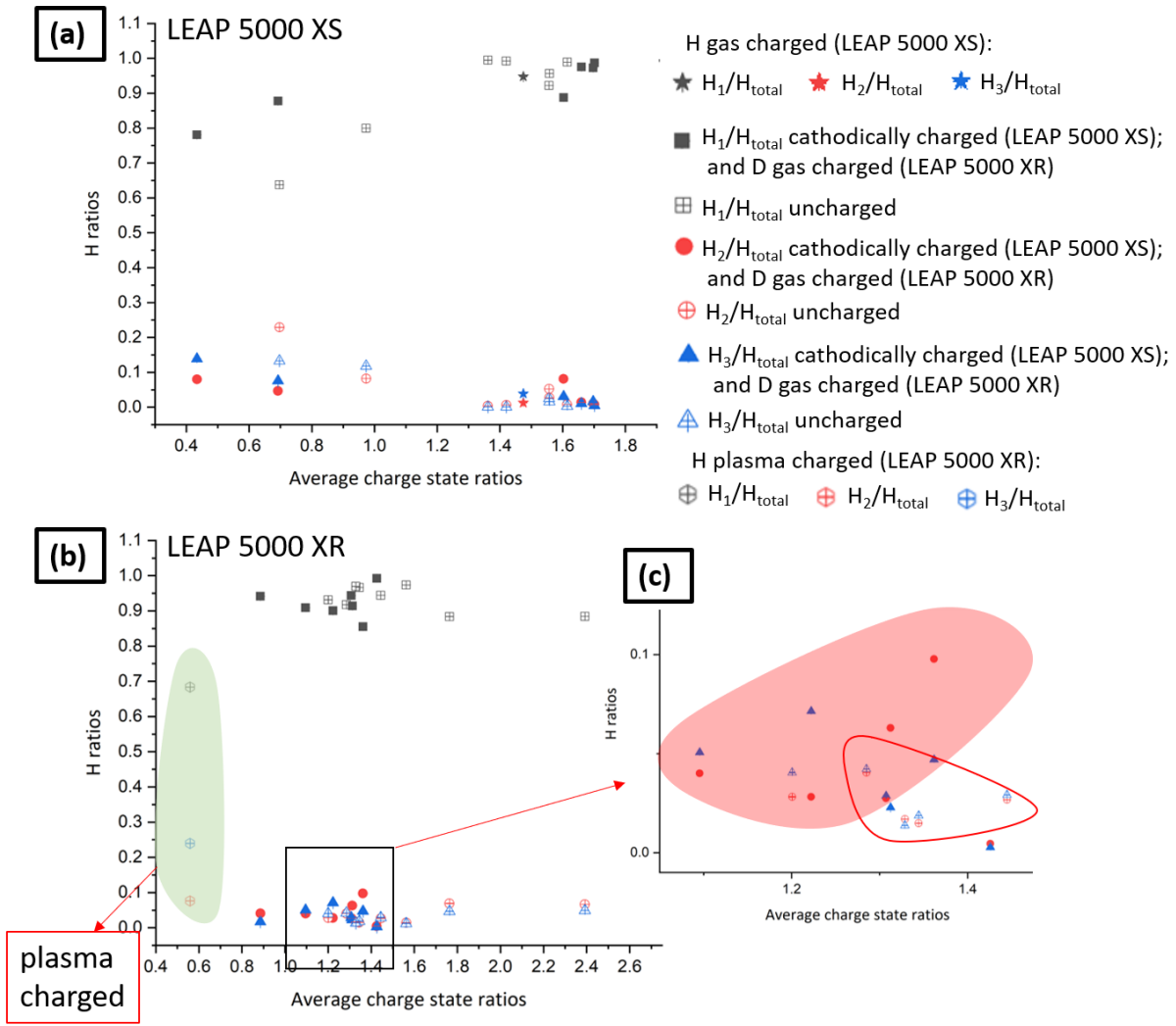

*Figure 12. (a) Average charge state ratios (CSR) vs. relative H abundances for the experiments performed in LEAP 5000 XS; (b) the average CSR vs. relative H abundances for the experiments performed in LEAP 5000 XR; (c) enlarged region delineated by the black box in Figure 12 (b).*

Figure 12 (c) is a close-up on a section of charge-state ratios. The region highlighted in red qualitatively illustrates the region with higher relative $H_2$ and $H_3$ contents than the region delineated by the red line that contains mostly data from uncharged specimens. The amount detected remains relatively low but appears significantly higher than the background and above the level in the uncharged specimens. It has been reported that higher loading fugacities are achieved using cathodic charging than plasma charging, which is higher than low pressure gas charging [2,49]. However, gas charging by using the RHM is the cleanest approach for charging APT specimens prepared by site-specific lift-out. The moisture content encountered in the mass spectra of specimens subjected to gas charging was negligible and the transfer of specimens to the PFIB for cleaning was also not required after gas charging.

We could later explore the possibility to equip the RHM with a plasma generator for instance, or make use of the one directly on the commercial LEAPs, which would help maintain the cleaner, UHV-chain. This would enable more facile analyses of plasma-charged specimens, through a cleaner approach compared to electrolytic charging, with quicker ingress kinetics than gas charging while allowing the pre- and post- charging data acquisition. However, radiation damage from the incoming energetic ions may also cause additional modification of the microstructure that should not be disregarded.

# 5 Conclusion

We have described three hydrogen/deuterium charging routes for charging site-specific APT specimens in the present study: cathodic charging, plasma charging and gas charging. A detailed step by step description of the corresponding workflows involved in optimizing the three charging routes is also given. The merits and drawbacks of the three charging methods are illustrated in the present study, while we demonstrate the success of a cathodic charging workflow and the gas charging approach. Although higher ingress kinetics may be achieved by plasma charging, but it may also damage and introduce defects into the microstructure. The origin of hydrogen/deuterium can still be questioned in case of cathodic charging due to its possible introduction during (re)sharpening of the specimen in the PFIB. Hence, we infer that the gas charging in the RHM provides the cleanest amongst our results, i.e. with least contamination, and would be the selected route to aim for quantification of hydrogen/deuterium in the considered material.

# 6 Acknowledgements

The authors would like to thank Uwe Tezins, Andreas Strum and Christian Bross for their support to the FIB, APT, UHV suitcase, Reacthub Module and $N_2$ glovebox facilities at MPIE. Ms. Monika Nellessen and Ms. Katja Angenendt are acknowledged for their support in SEMs and metallography sample preparation labs. H.K., S-H.K., L.T.S. and B.G. acknowledge the financial support from the ERC-CoG-SHINE-771602. R.S.V acknowledges the financial support from the IMPRS SurMat scholarship.